# Predicting Big Bang Deuterium

N. Hata, R.J. Scherrer, G. Steigman, D. Thomas and T.P. Walker[1]

*Department of Physics*

*The Ohio State University*

*Columbus, OH 43210*

## Abstract

We present new upper and lower bounds to the primordial abundances of deuterium and helium-3 based on observational data from the solar system and the interstellar medium. Independent of any model for the primordial production of the elements we find (at the 95% C.L.): $1.5 \times 10^{-5} \leq (D/H)_P \leq 10.0 \times 10^{-5}$ and $(^3He/H)_P \leq 2.6 \times 10^{-5}$. When combined with the predictions of standard big bang nucleosynthesis, these constraints lead to a 95% C.L. bound on the primordial abundance of deuterium: $(D/H)_{best} = (3.5^{+2.7}_{-1.8}) \times 10^{-5}$. Measurements of deuterium absorption in the spectra of high redshift QSOs will directly test this prediction. The implications of this prediction for the primordial abundances of helium-4 and lithium-7 are discussed, as well as those for the universal density of baryons.

---

[1] DOE Outstanding Junior Investigator

# Introduction

Of the primordially produced light elements D, $^3$He, $^4$He, and $^7$Li, the predicted abundance of D is the most sensitive to the baryon density (technically the baryon to photon ratio at the time of nucleosynthesis) and thus an accurate determination of the primordial D abundance serves, via standard big bang nucleosynthesis (BBN), as a sensitive diagnostic of the baryon density of the universe. Conversely, the primordial abundance of D can be used as a probe of the BBN model - any D abundance corresponds to definite BBN predictions for the abundances of $^3$He, $^4$He, and $^7$Li which can be tested with observational data. Unfortunately, D is the most weakly bound of the stable nuclei and is therefore easily destroyed in astrophysical environments (Epstein, Lattimer and Schramm 1976). This makes the extraction of the primordial D abundance from observations of the present interstellar medium (ISM) and from the solar system a complicated and, necessarily, model-dependent task. Initial studies (Yang *et al.* 1984 (YTSSO); Dearborn, Schramm and Steigman 1986 (DSS); Walker *et al.* 1991 (WSSOK); Steigman and Tosi 1992 (ST92)) focused on the evolution of D plus $^3$He in an attempt to soften the dependence on chemical evolution. Recently Steigman and Tosi (1995(ST95)) improved on this earlier work by introducing a generic parameterization of D and $^3$He chemical evolution. Here we show that the ST95 analysis leads to a bound on the primordial abundances of D and $^3$He independent of the mechanism by which these elements are primordially produced.

Our constraints, when combined with the BBN D and $^3$He predictions, lead to a prediction for the primordial abundance of D which best satisfies the observational data from the ISM and solar system *and* the BBN predictions: $(D/H)_{best} = (3.5^{+2.7}_{-1.8}) \times 10^{-5}$ (unless otherwise noted, quoted errors throughout correspond to 95% C.L.). This predicted abundance is roughly an order of magnitude less than that inferred from the possible detection of D in a QSO absorption line system by Songaila *et al.* (1994) and Carswell *et al.* (1994) but is in agreement with a possible D detection (Tytler 1994) in a different QSO absorption line system. For this narrow range of best fit values for primordial D, BBN predicts a $^4$He mass fraction of $Y_{best} = 0.247 \pm 0.004$ and $(^7Li/H)_{best} = (2.9^{+4.4}_{-1.5}) \times 10^{-10}$. The predicted $^4$He abundance does not agree with the primordial abundance as inferred from metal poor HII regions, possibly indicating an underestimate of the systematic errors in that data set. The predicted $^7$Li abundance is consistent with the data from metal-poor stars in our Galaxy's halo. The baryon density (relative to the critical density) consistent with this D prediction is $\Omega_B h_{50}{}^2 = (0.07^{+0.04}_{-0.02})$, which is roughly a factor of two larger than earlier BBN estimates.

## The Observational Bounds on Primordial D and $^3$He

The determination of the primordial D abundance from present day observations of the ISM and solar system clearly requires knowledge of the chemical evolution history of the Galaxy - the D abundance decreases with time since D is destroyed in astrophysical environments. Using the argument that some of the destroyed D ends up as $^3$He and some $^3$He survives stellar processing, past studies have focused on an upper bound to the abundance of primordial D *plus* $^3$He in an attempt to reduce the chemical evolution dependence. If the post big bang production of $^3$He is neglected, the dependence on chemical evolution is characterized by a single parameter, $g_3$, the fraction of $^3$He that survives stellar processing. The amount of $^3$He which survives stellar processing depends on stellar models, the initial mass function and the history of star formation, and therefore $g_3$ hides all of our ignorance about the chemical evolution history of the Galaxy. Recently, Steigman and Tosi (1995)(ST95) provided a generic analytic framework for the evolution of D and $^3$He. Starting from this analysis we show below that it is possible to bound simultaneously both the primordial abundance of D and that of $^3$He. Any model which predicts the primordial abundances of D and $^3$He is subject to this constraint. Comparing our new constraints to the predictions of BBN permit us to predict accurately the best fit primordial D abundance.

In order to trace the evolution of the D abundance we follow ST95 and track the fraction of gas, $f(t)$, which never undergoes stellar processing by the time $t$. Since any D cycled through stars is destroyed, $f$ and the D survival factor, $f_2$, are the same: $f_2(t) \equiv X_2(t)/X_{2\,P} = f(t)$ (we use $X_i$ to indicate mass fractions and the subscript "P" denotes the primordial abundance). Again following ST95, the mass fraction of $^3$He at any time $t$ is bounded by: $X_3(t) \geq X_{3\,P} f(t) + (1-f(t))g_3(X_{3\,P} + 3X_{2\,P}/2)$, where $g_3$ is the fraction of $^3$He which survives stellar processing. Here the first term on the right hand side is unprocessed $^3$He, the second term is the amount of $^3$He (both primordial and secondary from primordial D burning) which survives stellar processing, and the inequality accounts for possible stellar $^3$He production.[1] We can combine these expressions describing D and

$$y_{2P} \leq \left(\frac{X}{X_P}\right) \left\{ \left[y_{23} + \left(\frac{1}{g_3} - 1\right) y_3\right] - \frac{1}{g_3}\left(\frac{y_{3P}}{y_{23P}} y_{23}\right) \right\}, \qquad (1)$$

where $X$ is the hydrogen mass fraction ($X/X_P = 0.92 \pm 0.08$ (ST95)) and subscript "23" denotes D+$^3$He. This is the equation discussed by ST95 and is an improved (due to the last term which corrects for primordial $^3$He) version of that used by YTSSO and subsequent analyses (DSS; WSSOK). The chemical evolution model dependence involved in this constraint lies entirely in the choice of $g_3$, the $^3$He survival fraction. ST95 examined the bounds to the baryon-to-photon ratio which result from eq.(1) if the BBN predictions for the primordial components are adopted. In order to constrain simultaneously primordial D and $^3$He, we may write eq.(1) as a quadratic in $y_{2P}$:

$$y_{2P}^2 + (y_{3P} - \frac{X}{X_P} y_2 - \frac{1}{g_3}\frac{X}{X_P} y_3) y_{2P} + \frac{X}{X_P}\left(\frac{1}{g_3} - 1\right) y_2 y_{3P} \leq 0. \qquad (2)$$

---

[1] Recently, Olive *et al.* (1994) and Tosi *et al.* (1994) have examined models including net stellar production of $^3$He.

$^3$He evolution to write a single inequality relating the the primordial abundances of D and $^3$He ($y_{2P}$ and $y_{3P}$, the number fractions relative to hydrogen, respectively) to their solar system abundances ($y_2$ and $y_3$, respectively):

Provided that $y_{3P}$ is not too large, there are two real solutions corresponding to the equality of eq.(2) which represent upper and lower bounds to $y_{2P}$ consistent with the solar system abundances of D and $^3$He and some chemical evolution model for the stellar processing of $^3$He. In addition we require that any particular allowed choice of $y_{2P}$ be greater than both the ISM and solar system abundances of deuterium (since deuterium is always destroyed in astrophysical environments, $y_{2P} \geq y_{2\odot} \geq y_{2ISM}$ where $y_{2ISM} = (1.6 \pm 0.4) \times 10^{-5}$ from the combined Copernicus and IUE (McCullough 1992) and HST (Linsky $et\ al.$ 1993) observations of the interstellar medium D/H ratio and $y_{2\odot} = (2.6 \pm 1.8) \times 10^{-5}$ is the solar system abundance of D(ST95)). In Fig.(1) we show the solution to eq.(2) in the $y_{2P}$ - $y_{3P}$ plane using Geiss' (1993) data for the solar system abundances of D and $^3$He and $Y = 0.28 \pm 0.04$ for the solar $^4$He mass fraction (see ST95 and Thomas $et\ al.$ (1995) for details). Following YTSSO and WSSOK, we adopt $g_3 \geq 0.25$ and show the 95% C.L. bound for $g_3 = 0.25$ as a solid curve, the interior of which represents primordial abundances of D and $^3$He consistent with the solar system abundances of D and $^3$He. The contour is obtained from a joint likelihood function which includes gaussian probability distributions for all observed quantities and the assumption that for any particular choice of allowed $y_{2P}$ all values of $y_{3P}$ are equally likely (that is, each point interior to our contour can be associated with some non-zero stellar $^3$He production, each value of which we consider equally likely since the ST95 parameterization carries no information regarding stellar $^3$He production)[2]. We also incorporate a rigorous statistical definition in treating upper/lower bounds [for a complete discussion of the primordial abundance error analysis, see Thomas *et al.* (1995) and Hata *et al.* (1995)].

For comparison, YTSSO, DSS, and WSSOK adopted $g_3 \geq 0.25$ and ST92's results (as reanalyzed by ST95) show $g_3 \approx 0.6$. We feel our choice for $g_3$ is conservative with respect to most chemical evolution estimates for $g_3$ in the sense that recent studies find larger $g_3$ than assumed here (see, for example, Olive *et al.* (1994)) and the fact for any star $g_3 \geq 0.25$ (DSS). The trend is that the larger $g_3$, the more restrictive our constraint. For example, if we choose $g_3 = 0.6$ the bound on D for a given $^3$He abundance decreases by $\sim 40\%$. We could also take $g_3$ to be distributed between 0.25 and 0.5 with equal probablility. In this case, the upper bound to D for a given $^3$He abundance decreases by $\sim 30\%$. In contrast, $g_3$ could be smaller than 0.25 if a significant fraction of material has been cycled through two or more generations of stars. We have also plotted the bound from WSSOK on the primordial abundance of D+$^3$He: $y_{23\,P} \leq 12.5 \times 10^{-5}$ (this is the $2\sigma$ bound recalculated with the new solar system abundances of Geiss(1993)). Our upper bound to primordial D is more restrictive than WSSOK due to our more rigorous treatment of the upper and lower bounds. Independent of any model for primordial D and $^3$He production and assuming chemical evolution consistent with $g_3 \geq 0.25$ we find,

$$1.5 \times 10^{-5} \leq \left(\frac{D}{H}\right)_P \leq 10.0 \times 10^{-5} \quad \text{and} \quad \left(\frac{^3He}{H}\right)_P \leq 2.6 \times 10^{-5}. \tag{3}$$

### The Best Fit D and $^3$He Abundances

The key to determining the best fit value of the primordial D abundance is to compare the BBN predicted D vs. $^3$He relation with our D vs. $^3$He bounds from solar system and ISM data. In Fig.(1), we show the results of a Monte Carlo simulation of the standard[3] BBN reaction network. The dotted lines show the $1\sigma$ bounds to the BBN predictions for D and $^3$He. The numbers interior to this band (3, 4, 5, ...) represent the values of $\eta_{10}(\equiv \eta \times 10^{10}$, where $\eta$ is the baryon-to-photon ratio) corresponding to abundances of D and $^3$He predicted by standard BBN. Over the range of D and $^3$He abundances of interest the correlation of the errors in each abundance do not appreciably affect the BBN D vs. $^3$He relation and we construct the $1\sigma$ contour with the usual $\chi^2$ analysis. Our BBN code is an updated version of the Wagoner code (Wagoner 1973) with the errors in the reaction rates as detailed in Smith, Kawano, and Malaney (1993) (the exception being the neutron lifetime which we take to be $\tau_n = 887 \pm 2$ s in accordance with the latest PDG (1994)) and the various corrections to $Y_{BBN}$ as given by Kernan (1993). For further details, see Thomas *et al.* (1995) and Hata *et al.* (1995). Since D is burned more easily to $^3$He and $^3$H than $^3$H and $^3$He are burned to $^4$He, $^3$He decreases less rapidly with $\eta$ than does D and the $y_{2\,BBN}$ vs. $y_{3\,BBN}$ relation is monotonically increasing as seen in Fig.(1). Recall that high values of $y_{2\,BBN}$ and $y_{3\,BBN}$ correspond to low values of $\eta$ and, vice-versa, so that in Fig.(1) $\eta$ increases from upper right towards lower left.

---

[2] We thank M. Turner for two useful criticisms of the original version of this paper: (1) we included only the ISM lower bound to $y_{2\,P}$ and (2) our original normalization scheme for the likelihood function assumed that each value of $y_{3\,P}$ was equally likely and thus over-weighted the larger values of $y_{3\,P}$. The first correction increases the lower bound to D/H while the second correction only slightly alters the allowed region.

As seen in Fig.(1), the constraint on D and $^3$He can be significantly narrowed when our model-independent constraint is combined with the BBN prediction. To determine the best fit abundances from a joint fit, we construct a likelihood function including both the observational distributions and the probability distributions of the BBN predictions. The combined fit region (shown shaded in Fig.(1)) is obtained from the $\Delta\chi^2$ method (PDG 1994), where $\chi^2 = -2\log\mathcal{L}$ and $\mathcal{L}$ is our likelihood function.[4] This contour represents the best fit of the standard BBN predictions to our observational constraints. It corresponds to:
$$\left(\frac{D}{H}\right)_{best} = (3.5^{+2.7}_{-1.8}) \times 10^{-5} \quad \text{and} \quad \left(\frac{^3He}{H}\right)_{best} = (1.2 \pm 0.3) \times 10^{-5}. \qquad (4)$$

### Discussion

A clear confrontation between these BBN D predictions and anticipated observations of D in the spectra of QSO absorbers lies on the horizon. Recently Carswell *et al.* (1994)

---

[3] By "standard" we mean that the universe is assumed homogeneous and isotropic and that there are 3 light neutrino families.

[4] The $\Delta\chi^2$ method is justified when the likelihood function is a gaussian. Our likelihood function is well approximated by a gaussian in $(y_{2P}, y_{3P}, \eta)$.

and Songaila *et al.* (1994) reported $(D/H)_{QSO} \sim (1.9 - 2.5) \times 10^{-4}$ for a particular line of sight. This is roughly an order of magnitude larger than our prediction and at face value appears problematic for BBN and/or our treatment of chemical evolution. However, single QSO absorption system measurements may only provide upper limits to the D/H ratio due to the possibility of hydrogen interlopers masquerading as deuterium absorbers (Carswell *et al.* 1994; Songaila *et al.* 1994; Steigman 1994). In addition, reconciling such high D abundances with the ISM deuterium abundance and solar system $^3$He abundance proves difficult (Steigman 1994; Vangioni-Flam and Casse 1994; Olive *et al.* 1994). Recently, Tytler (1994) has reported a possible detection of D in a high redshift absorber at a level of $\sim 2 \times 10^{-5}$. While this apparently excellent agreement with our prediction could be viewed as a dramatic confirmation of standard BBN, it may indicate the natural dispersion of the inferred D abundances due to both hydrogen interlopers masquerading as deuterium and actual deuterium absorption at a level reduced by astration. Although it is clear that the QSO approach will provide an opportunity for direct measurement of the primordial D abundance, we caution the reader against drawing strong conclusions from the current statistically limited sample.

There are indirect tests of the consistency our best fit deuterium abundance. For our D constraint, the corresponding best fit value of $\eta$, $\eta_{best} = (5.0^{+2.9}_{-1.5}) \times 10^{-10}$, is larger (by roughly a factor of two) than that which results from previous analyses(see, for example, WSSOK) which used $^7$Li along with D+$^3$He to delineate the acceptable range of $\eta$. We now examine the predictions of the other primordial light elements consistent with our best fit D abundance. From the data of Fig.(1), the predicted abundances corresponding to $(D/H)_{best} = (3.5^{+2.7}_{-1.8}) \times 10^{-5}$ are:

$$\left(\frac{^3He}{H}\right)_{best} = (1.2 \pm 0.3) \times 10^{-5}, \tag{5}$$

$$Y_{best} = 0.247 \pm 0.004, \tag{6}$$

and

$$\left(\frac{^7Li}{H}\right)_{best} = (2.9^{+4.4}_{-1.5}) \times 10^{-10}. \tag{7}$$

The $^3$He prediction is currently the least testable of the three predictions because the chemical evolution of $^3$He is so poorly understood. This prediction is consistent with the smallest abundances of $^3$He inferred from observations of Galactic HII regions (Balser *et* *al.* 1994) and supports the idea that the abundance of $^3$He increases as the Galaxy evolves even if net stellar production is neglected (ST95).

The $^7$Li prediction is in good agreement with the primordial abundance of $^7$Li as inferred from abundances measured (see for example, Spite and Spite (1993) or Thorburn (1994)) in Pop. II halo stars : $(^7Li/H) \approx 1.2 \times 10^{-10}$ is the weighted average of the Li abundances for stars in the "Spite plateau" (WSSOK). The extraction of the primordial $^7$Li abundance from this data is model-dependent as the competition between stellar surface depletion and creation by cosmic ray nucleosynthesis must be taken into account. The standard depletion models (Deliyannis, Demarque and Kawaler 1990 and references therein) and the recent work of Vauclair and Charbonnel (1994) are consistent with modest depletion and an inferred primordial $^7$Li abundance of $(^7Li/H)_P = (1.2^{+4.0}_{-0.5}) \times 10^{-10}$ (the errors take into account possible production and depletion (see for example Copi, Schramm and Turner (1995)), in good agreement with our best fit value (see Thomas *et al.* (1995) for further discussion).

The largest discrepancy between the best fit predictions and observations occurs for $^4$He. This is not a new problem for BBN, but it becomes worse when one uses the predicted $^4$He abundance consistent with our best fit primordial deuterium result. Like all the other primordial elements, the primordial abundance of $^4$He must be inferred from data contaminated by stellar processing. Linear extrapolation of the $^4$He abundance in metal poor HII regions to zero metallicity yields (Olive and Steigman 1995): $Y_P = 0.232 \pm 0.003 \pm \Delta Y_{sys}$ where 0.003 is the $1\sigma$ statistical error and $\Delta Y_{sys}$ represents all the systematic uncertainties in converting $^4$He$^+$ line strengths into $^4$He mass fractions. Estimates for $\Delta Y_{sys}$ range from 0.005 (WSSOK; Pagel 1993; Olive and Steigman 1994), to 0.015 (Copi, Schramm and Turner (1995)). Our best fit prediction for the abundance of $^4$He disagrees with the observational data unless *the systematic errors are large (~ 0.015)*. If the systematic errors should not prove this large, then our best fit deuterium abundance challenges the consistency of the standard BBN model.[5] Such a potential conflict raises an interesting question: what is the best way to test standard BBN? We believe the answer crucially depends upon which of the primordial light elements can be best measured, both in terms of the quality of the observational data *and* the model dependence of the extrapolation from non-primordial to primordial abundances. From our viewpoint, there are several choices: (1) use the QSO absorption system technique (or perhaps D absorption in the halo of our Galaxy and neighboring galaxies) to establish a nearly primordial D abundance, (2) reduce the systematic errors on measurements of $Y$ in metal-poor HII regions, or (3) refine stellar models for $^7$Li depletion in Pop II stars. In option (1), the model dependence is small but the scatter in the data may be large. Both options (2) and (3) do not have the same baryon-diagnostic power (thru BBN) as does (1) and therefore the payoff from refining the data may not be as great. In two upcoming papers (Thomas *et al.* (1995) and Hata *et al.* (1995)) we discuss this in greater detail.

---

[5] As theorists we would be remiss in not pointing out several variations to the standard BBN model which can resolve this crisis by reducing the predicted abundance of $^4$He: (1) massive tau neutrinos (Kawasaki *et al.* (1993); Dodelson *et al.* (1993)), (2) degenerate neutrinos (Beaudet and Yahil (1977)), (3) short-lived decaying particles (Scherrer and Turner (1988)).

Putting aside for the moment the possible conflict between our predicted and the observed $^4$He, the best fit D abundance corresponds to a baryon to photon ratio of $\eta_{best} = (5.0^{+2.9}_{-1.5}) \times 10^{-10}$. The upper bound to $\eta$ is *model independent* in the sense that it relies only on measuring the ISM and pre-solar abundances of D. The lower bound is more model dependent since it is based on an extrapolation of the primordial abundances of D and $^3$He from solar system data. This range, higher by a factor of $\sim 2$ than that of WSSOK (due to our more restrictive constraint), corresponds to a baryon density parameter (for $T_{\gamma 0} = 2.726 \pm 0.010$ (Mather *et al.* 1990)):

$$\Omega_B h_{50}^2 = (0.07^{+0.04}_{-0.02}). \tag{8}$$

For $H_0 < 100$ km/s/Mpc, $\Omega_B \geq 0.013$ which reinforces the conclusion (WSSOK) that some, perhaps most, of the baryons in the Universe are "dark". Despite our higher range for $\eta$, the evidence for non-baryonic dark matter persists since for $H_0 > 40$ km/s/Mpc, $\Omega_B < 0.18$. This higher range for $\eta$ goes part of the way towards resolving the x-ray cluster baryon crisis (White *et al.* 1993; Steigman and Felten 1994). Steigman and Felten (1994) find that the current x-ray data suggest $\Omega h_{50}^{1/2} < 0.073 \eta_{10}$ which for our best fit $\eta$ yields the bound $\Omega h_{50}^{1/2} < 0.58$ and so the x-ray "crisis" still exists in the sense that for $\Omega = 1$, $H_0 < 17$ km/s/Mpc.

## Acknowledgements

We thank Paul Langacker for many useful conversations. We also thank Craig Copi, Dave Schramm, and Michael Turner for useful criticisms of the original version of this paper. This work is supported by the DOE (DE-AC02-76ER01545).

**Figure Caption**

**Figure 1.** Shown are observational bounds, BBN predictions, and the best fit values for the primordial abundances of D and $^3$He. The 95% C.L. bound (solid curve) on the primordial abundances of D and $^3$He (by number relative to hydrogen) assumes a $^3$He survival fraction $g_3 = 0.25$. Primordial abundances of D and $^3$He interior to this curve are consistent with the solar system abundances of D and $^3$He and the assumed chemical evolution parameterization. Also shown are the interstellar medium bound (dash-dot curve) and the bound from earlier work (WSSOK) (dashed curve). The $1\sigma$ range (dotted curves) of the BBN predictions for D vs. $^3$He (by number relative to hydrogen) are taken from our Monte Carlo of the BBN reaction network. The numbers interior to this band (3,4,5, ...) represent the values of $\eta_{10}$ (see text for definition) corresponding to the primordial D and $^3$He abundances predicted by standard BBN. The shaded region (dotted curve) represents the 68 (95)% C.L. region for the abundances of D and $^3$He obtained by a comparison of the constraints derived from solar system and ISM observations (solid curve and dash-dot curve) and the predictions of BBN (dotted band).

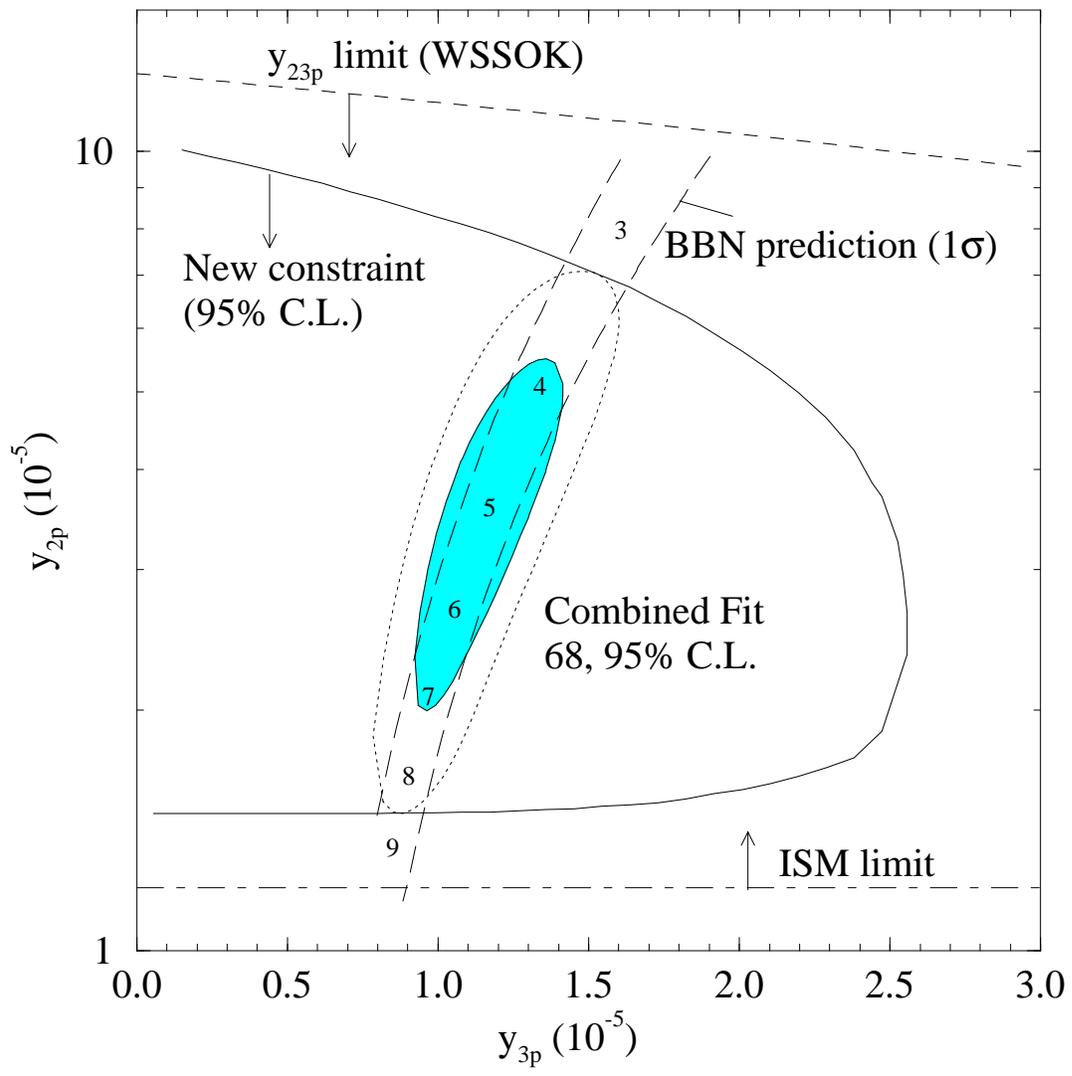